\documentclass{achemso}
\usepackage{graphicx}  
\usepackage{dcolumn}   
\usepackage{amssymb}
\usepackage{amsmath}
\usepackage{natbib}
\usepackage{mathrsfs}
\usepackage{color}
\usepackage{siunitx}
\usepackage{xr-hyper}
\usepackage{hyperref}
\mciteErrorOnUnknownfalse 

\usepackage[normalem]{ulem}

\title{Room temperature, cavity-free capacitive strong coupling to mechanical motion}

\author{Denise Puglia}
\altaffiliation{Equal contribution}
\email{denise.puglia@ist.ac.at} 
\affiliation{IST Austria, Am Campus 1, Klosterneuburg, 3400, Austria}

\author{Rachel Odessey}
\altaffiliation{Equal contribution}
\affiliation{IST Austria, Am Campus 1, Klosterneuburg, 3400, Austria}
\affiliation{Pritzker School of Molecular Engineering, University of Chicago,  5640 S Ellis Ave, Chicago, IL 60637, United States}

\author{Peter S. Burns}
\affiliation{IST Austria, Am Campus 1, Klosterneuburg, 3400, Austria}
\author{Niklas Luhmann}
\affiliation{Institute of Sensor and Actuator Systems, TU Wien, Gußhausstraße 27-29, Vienna, 1040, Austria}

\author{Silvan Schmid}
\affiliation{Institute of Sensor and Actuator Systems, TU Wien, Gußhausstraße 27-29, Vienna, 1040, Austria}

\author{Andrew P. Higginbotham}
\email{ahigginbotham@uchicago.edu}
\affiliation{James Franck Institute and Department of Physics, University of Chicago,  929 E 57th St, Chicago, Illinois 60637, USA}
\affiliation{IST Austria, Am Campus 1, Klosterneuburg, 3400, Austria}

\begin{document}

\begin{abstract}
The back-action damping of mechanical motion by electromagnetic radiation is typically overwhelmed by internal loss channels unless demanding experimental ingredients such as superconducting resonators, high-quality optical cavities, or large magnetic fields are employed.
Here we demonstrate the first room temperature, cavity-free, all-electric device where back-action damping exceeds internal loss, enabled by a mechanically compliant parallel-plate capacitor with a nanoscale plate separation and an aspect ratio exceeding 1,000.
The device has four orders of magnitude lower insertion loss than a comparable commercial quartz crystal and achieves a position imprecision rivaling optical interferometers.
With the help of a back-action isolation scheme, we observe radiative cooling of mechanical motion by a remote cryogenic load. 
This work provides a technologically accessible route to high-precision sensing, transduction, and signal processing.
\end{abstract}


Light incident upon a massive body exerts a famously weak optomechanical force \cite{lebedew_untersuchungen_1901,nichols_preliminary_1901,einstein_development_1909}.
For the particular case of an oscillating mass with average speed $v$, Braginskiĭ and Manukin argued that the optomechanical coupling is suppressed by a factor of $(v/c)^2$, and is therefore a small quantity due to the enormity of the speed of light $c$  \cite{braginski_ponderomotive_1967}.
Their realization that a cavity can resonantly enhance this coupling gave birth to the field of cavity optomechanics \cite{braginski_ponderomotive_1967,braginski_investigation_1970,aspelmeyer_cavity_2014}, in turn leading to experimental breakthroughs such as quantum ground-state cooling \cite{rocheleau_preparation_2010,teufel_sideband_2011,chan_laser_2011,rossi_measurement_2018,magrini_real-time_2021}, quantum transduction \cite{bochmann_nanomechanical_2013,andrews_bidirectional_2014,bagci_optical_2014,vainsencher_bidirectional_2016,higginbotham_harnessing_2018,mirhosseini_superconducting_2020,peairs_continuous_2020,delaney_superconducting-qubit_2022,sahu_entangling_2023, bozkurt_quantum_2023}, and detection of gravitational waves \cite{abramovici_ligo_1992,abbott_observation_2016}.

In the electrical domain, the parametric dependence of capacitance on position has long been understood as a source of optomechanical coupling \cite{ekinci_nanoelectromechanical_2005}.
Capacitive coupling is natively weak because of a mismatch, $Z_0/Z \ll 1$, between the small transmission-line impedance $Z_0$ and the large capacitive mechanical impedance $Z$ \cite{truitt_efficient_2007,ekinci_nanoelectromechanical_2005}.
The impedance mismatch, in fact, re-expresses the suppression noted by Braginskiĭ and Manukin; for a linear geometry $Z_0/Z \sim v/c$ (Supplement Sec.~\ref{sup-sec:coupling}).
The otherwise weak capacitive optomechanical interaction can be enhanced by application of a strong pump voltage \cite{truitt_efficient_2007}.
The particular case of a constant pump voltage generates linearized coupling between itinerant electromagnetic fields and mechanical motion.
As we show below, even in the absence of a cavity, reflected fields then exert a back-action force on mechanical motion, causing damping.
We refer to the regime where back-action damping exceeds all other dissipation as ``cavity-free capacitive strong coupling.''
Cavity-free capacitive strong coupling is analogous to the strong coupling regime of cavity optomechanics (Supplement Sec.~\ref{sup-sec:comparison}), to direct strong coupling of superconducting qubits to waveguides \cite{hoi_generation_2012,astafiev_resonance_2010,hoi_demonstration_2011,abdumalikov_dynamics_2011,garcia-ripoll_ultrastrong_2017,kannan_generating_2020}, and to over-coupling a resonant circuit to a transmission line.

Cavity-free capacitive devices based on silicon nitride membranes and graphene have been investigated previously \cite{xu_radio_2010,schmid_single_2014,mathew_dynamical_2016,schmid_fundamental_2016} and have so far fallen several orders of magnitude short of cavity-free capacitive strong coupling (see Supplement Sec.~\ref{sup-sec:lit_coop}).
Capacitive efficient \cite{truitt_efficient_2007} and strong \cite{bagci_optical_2014,bozkurt_quantum_2023} coupling has been achieved to high-impedance resonator (i.e. cavity) modes.
However, cavity-free strong coupling to itinerant modes, which have a low impedance bounded by $\sqrt{\mu_0/\epsilon_0} \approx 377~\mathrm{\Omega}$, has so far required the use of piezoelectric materials \cite{trolier_thin_2004,nguyen_mems_2007,kim_piezoelectric_2012,oconnell_quantum_2010,chu_quantum_2017,chu_creation_2018} or specialized experimental regimes involving high magnetic fields \cite{cleland_nanometre_1998,cleland_external_1999,cleland_nanomechanical_2002,knobel_nanometre_2003} or Coulomb blockades \cite{steele_strong_2009,lassagne_coupling_2009}.
No previous device to our knowledge has reached cavity-free capacitive strong coupling.

Here we reach cavity-free capacitive strong coupling, where electromechanical coupling to a $50~\mathrm{\Omega}$ environment overwhelms internal mechanical dissipation at room temperature without the use of a cavity.
We achieve 30,000 times ($45~\mathrm{dB}$) lower insertion loss and 10 times larger inductive bandwidth than a commercial piezoelectric device operating at similar frequencies.
We use this platform to efficiently detect mechanical motion with an imprecision under $20~\mathrm{fm/\sqrt{Hz}}$, rivaling optical Michelson interferometers \cite{pluchar_towards_2020, serra_silicon_2021,cupertino_centimeter_2024,bereyhi_perimeter_2022}.
Implementing a back-action isolation scheme, we verify that mechanical motion thermalizes to the external environment, and observe radiative cooling of mechanical motion by a remote cryogenic load.
Here a cryogenic load is required for cooling, unlike in cavity optomechanics where optical pumps naturally provide a cold bath for mechanical motion.

We identify two requirements for cavity-free capacitive strong coupling: \textit{(1)} that, by definition, the total electromechanical coupling rate, $2 g$ for a two-port device, exceed the internal dissipation rate $\kappa_\mathrm{in}$, and \textit{(2)}~that, in practice, no antiresonance obscure useful circuit properties.

\begin{figure}[b]
    \includegraphics[width=0.43\textwidth]{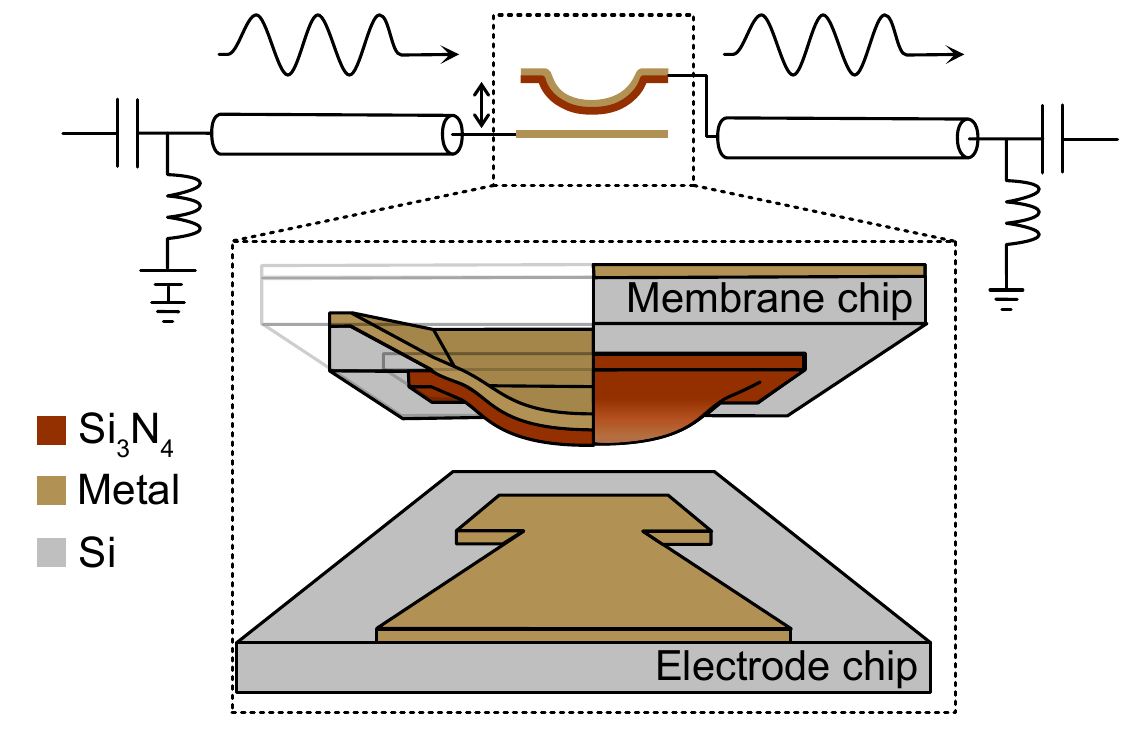}
    \caption{\label{fig:0} \textbf{Experimental setup.} A metallized $\mathrm{Si_3 N_4}$ membrane chip is flipped on an electrode chip and both are coupled to transmission lines, allowing the transmission $S_{21}$ to be measured.
    The front left quadrant of the membrane chip is omitted from the schematics and only the contour lines are shown.
    A DC bias tunes the coupling between the transmission lines and the flipped chip.  }
    \end{figure}
To achieve these requirements, we have devised a system consisting of a metallized $\mathrm{Si_3 N_4}$ membrane chip flipped on top of an electrode chip, with each chip coupled to a $50~\mathrm{\Omega}$ transmission line (Fig.~\ref{fig:0}).
The flipped pair of chips forms a movable capacitor $C(x)$ where $x$ is a coordinate parameterizing the mechanical displacement (see Supplement Section \ref{sup-part:theory}).
The electromechanical system is governed by the parametric interaction Hamiltonian $H_\mathrm{int} = -\frac{1}{2}C(x) V^2$, where $V=V_\mathrm{DC, S}+V_\mathrm{AC}$ is the combination of a large static DC and small oscillating AC voltage.
See Supplement Sec.~\ref{sup-sec:comparison} for a comparison between the parametric cavity-free capacitive and cavity optomechanical couplings.
The AC voltage results in an oscillating electromechanical force $F_\mathrm{em} = C' V_\mathrm{DC, S} V_\mathrm{AC}$ which is enhanced by $V_\mathrm{DC, S}$ (here $C'=\partial C/\partial x$ evaluated at equilibrium).
Decomposing the AC voltage into incoming and outgoing waves in the transmission line $V_\mathrm{AC}=V^+ + V^-$ reveals that there is a back-action contribution to $F_\mathrm{em}$ arising from the reflected wave.
The constitutive capacitive relation $q = C(x) V$ can be differentiated to show that this back-action force is phase delayed, generating electromechanical damping at a rate 
\begin{equation}
    \label{eq:g}
    g=C'^2V_\mathrm{DC,S}^2 Z_0 /m,
\end{equation}
where $Z_0$ is the transmission line impedance and $m$ is the motional mass of the membrane.
Although we find the distributed-element language convenient for describing our apparatus, we note that at our medium radio frequencies accumulated phases in the transmission lines are small, and the transmission line can be equivalently understood as a lumped-element resistor with resistance $Z_0$.

Requirement \textit{(1)}, $2 g / \kappa_\mathrm{in} > 1$, is analogous to reaching the high-cooperativity regime in cavity optomechanics and implies that dissipation is dominated by back-action damping from the environment (Supplement Sec.~\ref{sup-sec:comparison}).
We therefore introduce an analogous ``cooperativity'' metric, $\mathscr{C}=2 g / \kappa_\mathrm{in}$.
We achieve $\mathscr{C}>1$ by maximizing $C'$ with large typical membrane areas of $500\times500~\mathrm{\mu m^2}$ and small typical membrane-electrode separations of $300~\mathrm{nm}$.
Previous cavity-free capacitive devices have fallen short of $\mathscr{C}>1$ by many orders of magnitude, typically achieving $10^{-2} - 10^{-7}$ (see Supplement Section~\ref{sup-sec:lit_coop}).
Note that for a given geometry, the maximum achievable $V_\mathrm{DC}$ is limited by instability (see Supplement Section \ref{sup-sec:collapse_physics}).

Requirement \textit{(2)} for preserving useful circuit properties is motivated by the unavoidable presence of an antiresonance due to the electrostatic gate.
This is accomplished by detuning the mechanical resonant frequency $\Omega_m$ from the antiresonant frequency $\Omega_a$.
The mechanical resonance is shifted from its zero-voltage value $\Omega_0$ according to
\begin{equation}
    \label{eq:omega_m}
    \Omega_m^2=\Omega_\mathrm{0}^2 - \Omega_e^2,
\end{equation}
where $\Omega_e=\frac{C''V_\mathrm{DC, S}^2}{2m}$ and $C'' = \partial^2 C/\partial x^2$.
The antiresonance $\Omega_a^2$ is voltage-tuned according to 
\begin{equation}
    \label{eq:omega_a}
    \Omega_a^2 = \Omega_m^2 + \frac{g}{Z_0 C}.
\end{equation}
Equation~\ref{eq:omega_a} indicates that, in our approach, strong antiresonance detuning is an automatic consequence of having large $g$.
In fact, the antiresonance detuning $\Omega_a-\Omega_m \approx (g/2) (Z/Z_0)$ is controlled by $Z/Z_0$, which, as discussed above, is large in practice.
Strongly detuning the antiresonance gives unobscured access to large effective mechanical inductance, a desirable feature for electrical circuits.

\begin{figure}
\includegraphics[width=0.48\textwidth]{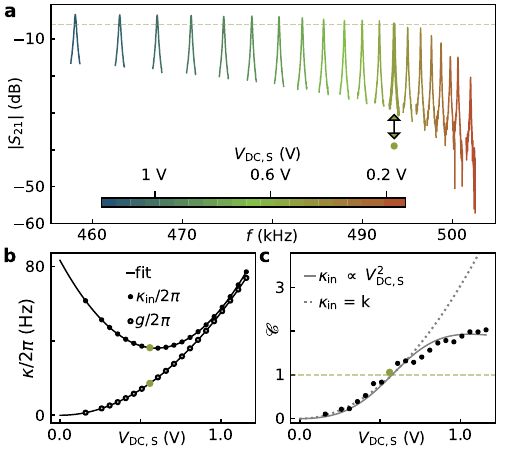}
\caption{\label{fig:1} \textbf{Tuning from weak to strong electromechanical regime.} 
\textbf{a}, $S_\mathrm{21}$ transmission parameter around the resonance frequency of the fundamental mode for different voltages. 
The dashed green line indicates expected threshold transmission when the device is operated in the cavity-free capacitive strong coupling regime $2g > \kappa_\mathrm{in}$. 
\textbf{b}, External coupling $g$ and internal coupling $\kappa_\mathrm{in}$ as a function of voltage.
Green marker was calculated from the highlighted trace in \textbf{a}. 
The full lines represent the voltage dependence fits. 
\textbf{c}, Cooperativity $\mathscr{C}$ as a function of voltage, with the green marker calculated from the highlighted trace in \textbf{a}. 
The dashed green line shows $\mathscr{C} = 1$, which separates capacitive cavity-free strong and weak coupling.
The full line indicates the expected cooperativity based on the voltage dependence fit from \textbf{b}, while the dotted line indicates the expected cooperativity for a constant internal coupling. }
\end{figure}

The strength of electromechanical coupling is experimentally assessed by measuring the transmission coefficient $S_\mathrm{21}$ around the fundamental mode for different DC bias voltages.
For $V_\mathrm{DC} = 0~\mathrm{V}$, a narrow, resonant peak in transmission is observed, suggesting weak but nonzero electromechanical coupling at low bias (Fig.~\ref{fig:1}a). 
Increasing $V_\mathrm{DC}$ causes the resonance to increase and widen, indicating a growth in electromechanical coupling.
For large voltage ($\gtrsim$ 0.4 V), the resonant transmission becomes voltage independent, saturating at a value slightly below unity.

The values of $g$ and $\kappa_\mathrm{in}$ were extracted from a Lorentzian fit of $S_\mathrm{21}$; $\kappa_\mathrm{in}$ includes both intrinsic mechanical loss and any parasitic resistive losses (see Supplement Section \ref{sup-sec:couplingextraction}) \cite{cleland_external_1999}.
Figure \ref{fig:1}b shows that $g$ depends quadratically on applied voltage over the entire measurement range, as qualitatively expected from Eq.~\ref{eq:g}.
However, we consistently find that external coupling is minimized at non-zero DC voltage, presumably due to interfacial charges, motivating the use of a shifted bias $V_\mathrm{DC,S}=V_\mathrm{DC}-V_0$, where $V_0$ is sample dependent.
The voltage offset is determined by fitting $g(V_\mathrm{DC,S})$ to Eq.~\ref{eq:g} with $V_\mathrm{DC,S}=V_\mathrm{DC}-V_0$.
Internal dissipation is non-monotonic in $V_\mathrm{DC,S}$, reaching a minimum near 0.4 V and growing for large voltage, eventually matching the growth in $g$, which is the origin of the saturation in resonant transmission at large voltages.
We speculate that the quadratic voltage-dependence of $\kappa_\mathrm{in}$ is due to parasitic inline resistance from the electrodes \cite{cleland_external_1999}.
The relative voltage shift between $\kappa_\mathrm{in}$ and $g$ is not understood.
The minimum internal dissipation is limited to around 36$\pm$8~Hz for this sample, but reaches values down to 5$\pm$2~Hz for other samples (see Supplement Section \ref{sup-part:param}).

The competition between coupling and dissipation is summarized with the ``cooperativity'' metric $\mathscr{C}=2 g/\kappa_\mathrm{in}$ (Fig. \ref{fig:1}c).
As voltage is increased $\mathscr{C}$ increases from a small value eventually crossing the $\mathscr{C}>1$ threshold for cavity-free capacitive strong coupling.
This signals that the light-matter interaction has entered into a regime where mechanical motion is damped primarily by radiation in the leads. 
To the best of our knowledge, this is the first reported device to reach cavity-free capacitive strong coupling between electromagnetic radiation and mechanical motion.

\begin{figure}
\includegraphics[width=0.48\textwidth]{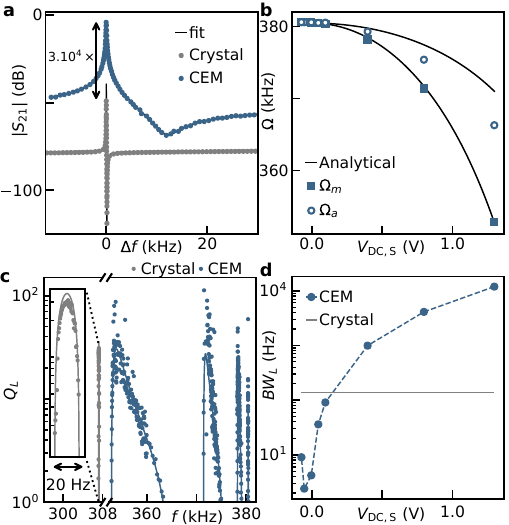}
\caption{\label{fig:2} \textbf{Comparison with a commercial device.} 
\textbf{a}, $S_\mathrm{21}$ around the resonance and antiresonance for the crystal and capacitive electromechanics (CEM) devices at $V_\mathrm{DC} = 1~\mathrm{V}$. Full line represents a fit. 
\textbf{b}, Antiresonance $\Omega_a$ and resonance $\Omega_m$ voltage dependence. The full line represents a fit of the voltage dependence from the mechanical resonance and expected antiresonance dependence based on this fit. 
\textbf{c}, Inductor quality factor $Q_L$ data (marker) and fit (full line). The inset shows the commercial quartz crystal (CC1V-T1A) resonance.
\textbf{d}, Inductive bandwidth $BW_L$, defined by the frequency range over which $\mathrm{Im}(Z) > 0$ for a fixed $V_\mathrm{DC,S}$ . Points are calculated from the data in \textbf{c}, and the line is a guide to the eye. Horizontal line indicates inductive bandwidth of the crystal.}
\end{figure} 

Cavity-free capacitive strong coupling opens up new avenues for electromechanical devices in regimes that are challenging for existing technology. 
To explore this, Fig. \ref{fig:2}a shows a comparison between the capacitive electromechanics (CEM) device and a commercial quartz crystal oscillator CC1V-T1A operating at a similar frequency.
CEM shows an insertion loss 30,000x (45 dB) lower than the crystal.
Both systems present an antiresonance as qualitatively expected from a Butterworth-van Dyke model \cite{horowitz_art_1994,truitt_efficient_2007}; however, we find that the CEM antiresonance is quantitatively not as deep as expected and requires an additional shunt resistance in the circuit model (see Supplement Section\ref{sup-sec:butterworth}). 
CEM also shows more than 12 kHz detuning between the resonance and the antiresonance, while the crystal has a detuning of less than a kHz.

Both the resonant and antiresonant frequencies are voltage tunable.
At low voltages, there is very little detuning of the antiresonance from the resonance. 
As voltage increases, the frequency of the resonance decreases faster than that of the antiresonance, increasing the detuning between the two (Fig.~\ref{fig:2}b).
Fitting the resonant frequency voltage dependence to Eq.~\ref{eq:omega_m} and allowing for a small voltage offset yields a zero-voltage frequency $\Omega_0$ is within 0.07\% of expectations based on fabrication parameters, and a zero-voltage separation distance $d=320~\mathrm{nm}$, within $20~\mathrm{nm}$ of the design value (see Supplement Section \ref{sup-sec:fabrication}).
The similarity between target parameters and fitted values demonstrates good fabrication control.
Equation~\ref{eq:omega_a} predicts the voltage dependence of the antiresonant frequency to within 1.2\% with no free parameters, although the experimental antiresonance systematically depends more strongly on voltage than expected.

Given the prevalence of quartz crystal oscillators in electrical engineering, it is natural to view CEM as a circuit element characterized by a lumped-element impedance $Z$.
Transforming the measured scattering parameters into an impedance reveals an inductive region ($Z \approx \mathrm{Im}(Z)>0$) at frequencies between the resonance and antiresonance (Fig.~\ref{fig:2}a).
We quantify the quality of this effective inductance with the metric $Q_L=\frac{\mathrm{Im}(Z)}{\mathrm{Re}(Z)}$ (see Supplement Section \ref{sup-sec:ql}), and find $Q_L\approx 30$, which is comparable to that of the commercial quartz crystal.
CEM maintains this $Q_L$ over more than 25 kHz of tunable bandwidth by adjusting the DC voltage.
At fixed DC voltage CEM has an instantaneous inductive bandwidth $BW_L$ surpassing $10~\mathrm{kHz}$, which is two orders of magnitude larger than the commercial crystal (Fig.~\ref{fig:2}d). 
Thus, CEM shows promise as a compact, low insertion loss, high $Q_L$, and wide band inductor for electrical engineering.

\begin{figure}
\includegraphics[width=0.48\textwidth]{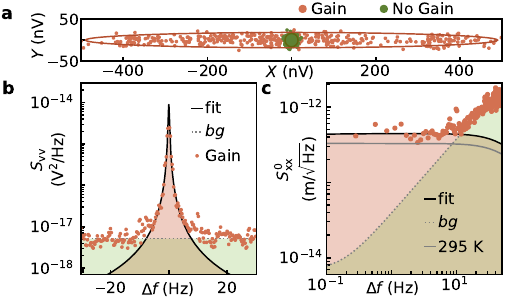}
\caption{\label{fig:3} \textbf{High-precision position measurements.} 
\textbf{a}, demodulated $X$ and $Y$ quadratures for CEM with (orange) and without (green) parametric gain. The full line represents 4x the standard deviation of the data.
\textbf{b}, the voltage spectral density $S_\mathrm{VV}$ for the parametrically amplified trace shown in \textbf{a}. 
The solid line is a fit to a Lorentzian with an offset $bg$ (dotted line). 
The green shading indicates the voltage noise of the measurement amplifier and the orange shading indicates the excess noise due to thermomechanical fluctuations. 
\textbf{c}, Position imprecision inferred from \textbf{b} using Eq.~\ref{eq:sxx}. The full gray line represents the expected contribution of thermomechanical fluctuations at 295 K.}
\end{figure}

A fundamental consequence of reaching strong electromechanical coupling ($\mathscr{C}>1$) is that mechanical motion thermalizes to radiation in the environment.
However, at our low operating frequencies, thermomechanical fluctuations are obscured by technical noise associated with the measurement chain. 
To overwhelm this technical noise, we introduce a parametric drive at twice the mechanical resonant frequency while measuring demodulated ($X(t)$, $Y(t)$) quadratures at the mechanical resonant frequency.
As shown in Fig.~\ref{fig:3}a, the introduction of a parametric drive causes a substantial gain in the $X$-quadrature noise, vastly exceeding the technical noise of the measurement, and no significant change in the $Y$ quadrature, as expected for this pumping configuration (see Supplement Section \ref{sup-sec:gain_frequency})).
The amplified $X$-quadrature is strongly peaked at mechanical resonance, consistent with the expected Lorentzian profile of amplified mechanical noise (Fig.~\ref{fig:3}b).

In the high-gain limit, the voltage fluctuation can be directly converted into equivalent undriven position imprecision $S^0_\mathrm{xx}$ using the relation
\begin{equation}
    \label{eq:sxx}
    S_\mathrm{VV} = gm\Omega_m^2|G(f)|^2S^0_\mathrm{xx}, 
\end{equation}
where $G(f)$ is the frequency-dependent gain of the quadrature being measured (see Supplement Section \ref{sup-sec:gain_frequency})). 
Near mechanical resonance, $S^0_\mathrm{xx}$ is dominated by mechanical fluctuations, and we find that technical noise contributes less than $20~\mathrm{fm/\sqrt{Hz}}$ of position imprecision, which is comparable to state-of-the-art cavity-free optical interferometers working towards ground state cooling \cite{pluchar_towards_2020}.
However, the thermomechanical fluctuations associated with the Lorentzian fit are equivalent to a temperature of $568\pm15~\mathrm{K}$, far exceeding room temperature.
This value is compatible with independent measurements of the electrical back-action noise of the electronics in our measurement setup (see Supplement Section \ref{sup-sec:backaction})).
Thus, unlike optical interferometers, our measurement apparatus gives substantial back-action noise.

\begin{figure}
\includegraphics[width=0.48\textwidth]{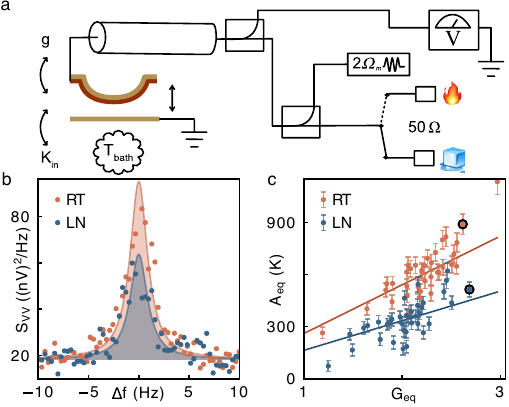}
\caption{\label{fig:4} \textbf{Back-action isolation.} \textbf{a}, CEM circuit in reflection and back-action isolation mode. The temperature of the external port is determined by a 50~$\Omega$ resistor at LN (ice cube) or room (fire) temperature. 
\textbf{b}, Voltage spectral density for an CEM device thermalized to RT and LN. The full line is a fit to a Lorentzian. The thermomechanical fluctuations are proportional to the area between the trace and the background, as indicated by the shading.
\textbf{c}, Lorentzian area $A_\mathrm{eq}$ as a function of gain $G_\mathrm{eq}$. 
Error bars are calculated based on the number of samples and the measurement bandwidth, as discussed in Supplement Sec. \ref{sup-sec:area}. 
The circle markers correspond to traces shown in \textbf{b}. 
The full lines represent fits of the gain dependence.}
\end{figure}
    
To address this problem, we have devised a simple back-action isolation scheme which protects the mechanical oscillator from measurement noise.
The input signal is added through the main line of a directional coupler and reflection is measured through the coupled port (Fig.~\ref{fig:4}a).
Attenuation from the coupled port reduces measurement back-action at the expense of increasing imprecision.
The main line carries voltage fluctuations from a reference $50~\mathrm{\Omega}$ resistor which, in the limit of strong electromechanical coupling, controls the equilibrium temperature of the mechanical resonator.
As a consequence, a cold reference resistor should radiatively cool mechanical motion.
Compared to the original transmission circuit configuration, the back-action isolation configuration attenuates output voltage fluctuations (\textit{cf.} Fig.~\ref{fig:4}b and Fig.~\ref{fig:3}a).
Cryogenically cooling the remote, $50~\mathrm{\Omega}$ load noticeably decreases the peak voltage spectral density, giving an initial indication that the mechanical temperature is lowered.
This observation can be quantified by calculating a temperature-equivalent area $A_\mathrm{eq}$ under the thermomechanical Lorentzian which measures the mechanical temperature $T_m$,
\begin{equation}
    \label{eq:area_vs_gain}
    A_\mathrm{eq} =  G_\mathrm{eq} T_m - \eta T,
\end{equation} 
where $\eta$ is the insertion loss from the membrane to the voltage measurement and $G_\mathrm{eq} = \eta G_A$ is the net mechanical gain. 
Fig. \ref{fig:4}c shows the numerically-integrated area as a function of $G_\mathrm{eq}$ and error bars given by the Dicke radiometer (see Supplement Section\ref{sup-sec:radiometer}).
The gain for each experimental run is independently measured with a pilot tone. 
Fitting to Eq.~\ref{eq:area_vs_gain} gives $T_m = 281\pm 7~\mathrm{K}$ for the room temperature load, a reasonable value.
The same procedure for the cold load yields $T_m = 171\pm 7~\mathrm{K}$, compatible with the expected value of $177~\mathrm{K}$ based on independently measured system and resonator losses (see Supplement Section \ref{sup-sec:tmech}).
This demonstrates the radiative cooling of a room-temperature mechanical resonator by a remote cryogenic load.

The spread of the points in Fig. \ref{fig:4}c is not compatible with statistical errors.
Rather, we believe it originates from gain instabilities during individual measurements.
Provided that these gain instabilities are cured, CEM could be a competitive 50 $\Omega$ matched amplifier and would achieve noise performance comparable to radio-to-optical receivers \cite{bagci_optical_2014}.
Alternatively, the CEM amplifier can be operated in high impedance environment, in which gain instabilities are not observed (see Supplement Section \ref{sup-sec:highimpedance}).

Summarizing, we have reported a platform to achieve cavity-free capacitive strong electromechanical coupling in an all-electric device. 
This voltage-tunable coupling reaches cooperativity $\mathscr{C} > 1$ for $V_\mathrm{DC}$ as low as $0.5~\mathrm{V}$.
From a circuit element perspective, CEM presents a region of inductive behavior that is wider and less lossy than in commercial quartz crystals.
These features could be useful in applications that require large, low-loss inductance, such as a compact step-up transformer for noise matching to transistor amplifiers.
We further demonstrate parametric amplification of the CEM and resolve thermomechanical fluctuations.
Amplifications result in a low equivalent position imprecision $S^0_\mathrm{xx}<20~\mathrm{fm/\sqrt{Hz}}$, suggesting future applications in sensing.
In this configuration, mechanical motion was heated by electrical back-action by more than 300 K.
A back-action isolation scheme removed the excess heating, enabling the cooling of mechanical motion to 171 K with a remote cryogenic resistor.
In addition to the possible applications mentioned above, we expect our approach to be generally useful in environments that are challenging for optical cavities or superconducting resonators.

To more fully compare with previous work, we have have estimated achievable $\mathscr{C}$ in common device geometries appearing throughout the literature, including based on two-dimensional materials (Supplement Sec.~\ref{sup-sec:lit_coop}).
Typical device geometries project to $\mathscr{C}$ in the range $10^{-2}-10^{-9}$ \cite{truitt_efficient_2007,mathew_dynamical_2016,xu_radio_2010,schmid_single_2014,zhou_high-q_2021}.
In contrast, cavity-coupled devices that can be sideband cooled nearly to the ground state project to reach $\mathscr{C}>1$ \cite{yuan_large_2015,andrews_quantum-enabled_2015}, suggesting that a select few electromechanical geometries could exhibit cavity-free capacitive strong coupling in the future.

Another useful point of comparison is Bagci et al. \cite{bagci_optical_2014}, which realizes radio-to-optical transduction via strong coupling between mechanical motion and a room temperature, high-impedance, radio-frequency resonator.
Loss from the room-temperature resonator limited performance.
Our approach, in contrast, does away with the lossy intermediate resonator, thereby presenting a possible route to improving overall performance.

Supporting information includes Ref.'s \cite{jacobs_quantum_2014,leissa_vibration_1969,rugar_mechanical_1991,bothner_cavity_2020,chen_graphene_2013}.

\textbf{Supporting information.} Fundamental coupling considerations, comparison with cavity optomechanics, further experimental details, theoretical model.

\section*{Acknowledgements}
We thank Carissa Kumar and Vibha Padmanabhan for assistance in comparing performance with devices across the literature.
We thank Andrew Cleland for helpful comments on this work.
We are grateful for support from the Miba Machine Shop and Nanofabrication facility at IST Austria.
This work was supported by the Austrian FWF grant P33692-N and includes a recipient of a DOC Fellowship of the Austrian Academy of Sciences (DOC – No. 26088) at the Institute of Science and Technology, Austria.


\providecommand{\latin}[1]{#1}
\makeatletter
\providecommand{\doi}
  {\begingroup\let\do\@makeother\dospecials
  \catcode`\{=1 \catcode`\}=2 \doi@aux}
\providecommand{\doi@aux}[1]{\endgroup\texttt{#1}}
\makeatother
\providecommand*\mcitethebibliography{\thebibliography}
\csname @ifundefined\endcsname{endmcitethebibliography}  {\let\endmcitethebibliography\endthebibliography}{}

\newcommand{\suppage}[1]{
	\pagebreak
	\begin{figure}[p]
		\vspace*{-1.5cm}
		\hspace*{-2.2cm}
		\includegraphics[page=#1,scale=0.95]{supplement/supplement.pdf}
	\end{figure}
}

\suppage{1}
\suppage{2}
\suppage{3}
\suppage{4}
\suppage{5}
\suppage{6}
\suppage{7}
\suppage{8}
\suppage{9}
\suppage{10}
\suppage{11}
\suppage{12}
\suppage{13}
\suppage{14}
\suppage{15}
\suppage{16}
\suppage{17}
\suppage{18}
\suppage{19}

\end{document}